\documentclass[10pt,osajnl,prl,twocolumn,showpacs,superscriptaddress,floatfix]{revtex4-1} 

\usepackage{amsmath}
\usepackage{graphics}
\usepackage{color}
\usepackage{caption}
\usepackage{subcaption}
\usepackage{float}
\usepackage{array}
\usepackage[margin=1.0in]{geometry}
\usepackage{ulem}
\usepackage{amssymb}

\begin{document}
\title{Self-Healing Organic-Dye-Based Random Lasers}
\author{Benjamin R. Anderson}
\author{Ray Gunawidjaja}
\author{Hergen Eilers}
\email{eilers@wsu.edu}
\affiliation{Applied Sciences Laboratory, Institute for Shock Physics, Washington State University,
Spokane, WA 99210-1495}
\date{\today}

\begin{abstract}
One of the primary difficulties in the implementation of organic-dye-based random lasers is the tendency of organic dyes to irreversibly photodecay.  In this letter we report the observation of ``reversible'' photodegradation in a Rhodamine 6G and ZrO$_2$ nanoparticle doped polyurethane random laser.  We find that during degradation the emission broadens, redshifts, and decreases in intensity.  After degradation the system is observed to self-heal leading to the emission returning to its pristine intensity, giving a recovery efficiency of 100\%.  While the peak intensity fully recovers, the process is not strictly ``reversible'' as the emission after recovery is still found to be broadened and redshifted.  The combination of the peak emission fully recovering and the broadening of the emission leads to a remarkable result: the random laser cycled through degradation and recovery has a greater integrated emission intensity than the pristine system.

\vspace{0.5em}

\end{abstract}

\maketitle
Random lasers are disordered laser materials in which scattering acts as the feedback mechanism leading to amplification.  These materials are promising for many applications including authentication \cite{Cao05.01,Anderson14.04}, biological imaging \cite{Redding11.01,Redding12.01}, and tunable light sources \cite{Cao05.01,Leonetti13.02}.  One of the primary classes of random lasing (RL) materials is based on organic laser dyes.  Organic laser dyes are attractive for RL as they are cheap and easy to produce.  However, one of the major limitations of using organic dyes in random lasers is the tendency of dyes to \textit{irreversibly} photodegrade under intense illumination \cite{Villar13.01,Sebastian14.01}.  Photodegradation of organic dyes has been an active area of research for some time with much progress in understanding the mechanisms of -- and mitigation of -- photodegradation \cite{Rabek95.01,wood03.01}. One of the major discoveries concerning the mitigation of photodegradation is the 
observation of \textit{reversible} 
photodegradation in some dye-doped polymers \cite{Peng98.01,howel02.01,zhu07.01, Anderson11.02, Hung12.01,Kobrin04.01}.  


Self-healing dye doped polymers were first observed in the late 1990's by Peng \textit{et al.} who were studying fluorescence in Rhodamine B and Pyrromethene dye-doped (poly)methyl-methacrylate (PMMA) optical fibers \cite{Peng98.01}.  They observed limited recovery ($\approx$ 20\%) of the fluorescence efficiency after photodegradation \cite{Peng98.01}.  Several years later Howell and Kuzyk observed fully reversible photodegradation in disperse orange 11 (DO11) dye-doped PMMA \cite{howel02.01}, after which self-healing was discovered in 8-hydroxyquinoline (Alq) doped PMMA \cite{Kobrin04.01}, air force 455 (AF455) doped PMMA \cite{zhu07.01}, various anthraquinone dye-doped PMMA \cite{Anderson11.02}, and a DO11 dye-doped copolymer of styrene and MMA \cite{Hung12.01}.  In all these reported cases of self-healing it is found that the dyes \textit{irreversibly} photodegrade when in liquid solution \cite{howel04.01} but display reversibility when doped into polymer \cite{howel02.01}, which suggests that the polymer 
plays a crucial role in reversibility \cite{Ramini12.01}. Additionally, it was found that for the same dye molecule the polymer composition affects the recovery characteristics \cite{Hung12.01}, once again suggesting the vital role of the polymer in reversible photodegradation.  Based on these observations we test the photodegradation of a Rhodamine 6G (R6G) dye-doped polyurethane (PU) random laser with ZrO$_2$ nanoparticles \cite{Anderson14.04} and observe ``reversible'' photodegradation with a recovery efficiency of 100\%.

\begin{figure}
\centering
\includegraphics{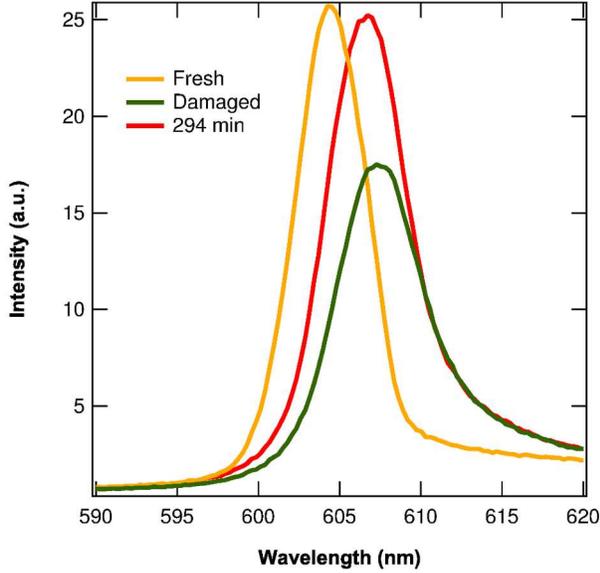}
\caption{Random laser emission spectra for the fresh sample, after being decayed for 21 min and after 273 min of recovery (294 min total).}
\label{fig:spectra}
\end{figure}

We fabricate the R6G+ZrO$_2$/PU random lasers by first dissolving R6G in solution of tetraethylene glycol and poly(hexamethylene diisocyanate), after which a dispersion of ZrO$_2$ in 1,4-dioxane and di-n-butyltin is added and the mixture is placed in a die to cure overnight \cite{Anderson14.04}.  Once cured, the samples are placed into a random lasing setup \cite{Anderson14.04} which focuses a 10 ns frequency doubled Nd:YAG laser into a spot of diameter 2.8 mm with a pulse energy of 33 mJ.  During photodegradation the sample is illuminated for 21 min and the RL emission is measured at 30 s intervals with the RL signal averaged over five pulses as the pump laser is found to have a shot-to-shot energy variation of 1\%. After photodegradation the pump beam is blocked and the sample is left in the dark except when taking recovery measurements, which consists of recording five-shot spectrum of the RL emission.  Figure \ref{fig:spectra} shows the emission intensity at the start of photodegradation (orange curve), 
after 21 min 
decay 
(green curve), and after 273 min of recovery (red curve).  From Figure \ref{fig:spectra} we find that after photodegradation the emission is less intense, red-shifted, and broadened.  Additionally, we see that after 273 min of recovery the emission is still red shifted, broadened, and the peak intensity is almost at the pristine level.

\begin{figure}
\centering
\includegraphics{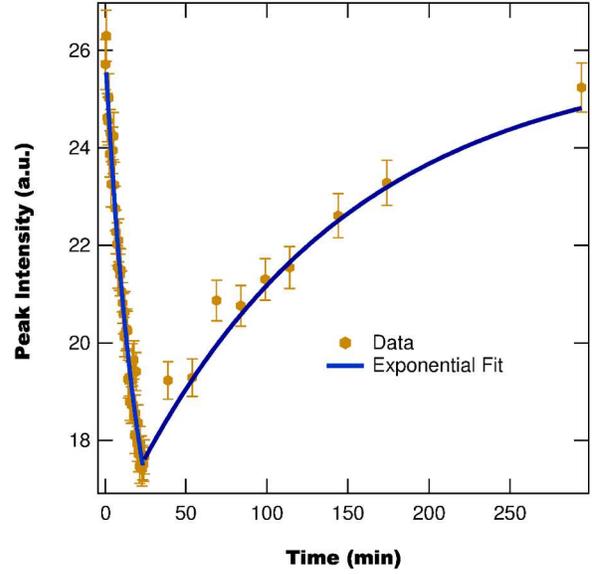}
\caption{Peak intensity as a function of time during decay and recovery.}
\label{fig:peak}
\end{figure}

To better quantify the degradation and recovery of the RL emission we consider the peak intensity, lasing wavelength, and linewidth during decay and recovery.  Figure \ref{fig:peak} shows the peak intensity as a function of time during decay and recovery.  The emission decays exponentially with a rate of $4.85(\pm 0.96) \times 10^{-2}$ min$^{-1}$ -- which converts to a photostability figure of merit (FoM) of $1.78(\pm 0.35)\times 10^{26}$ m$^{-2}$ with the FoM is defined as $I/\gamma$, where $\gamma$ is the decay rate and $I$ is the spatially averaged intensity in terms of photons/(s m$^2$) \cite{Gonzalez00.02}  -- and an asymptotic intensity (AI) of $13.4 \pm 1.2$. After the pump is turned off the sample begins recovering as an exponential function with a time constant of $137\pm 13$ min and an AI of $25.95 \pm 0.90$ which is within uncertainty of the pristine sample's emission intensity. This implies a recovery efficiency of 100\% for this level of pump fluence ($1.69$ kJ/cm$^2$).  However, 
experiments with larger fluences ($\approx 6.75$ kJ/cm$^2$) are found to have a recovery efficiency less than 100\% suggesting that the degree of recovery depends on the fluence during degradation. 

Additionally, preliminary decay and recovery cycling experiments -- in which the system is decayed at low fluence, allowed to fully recover, and then decayed again with the same pump fluence -- display full recovery after the second degradation cycle.  This suggests that for pump fluences below a certain value the R6G molecules degrade to a species which self-heals, while larger pump fluences result in the pristine R6G molecules either degrading directly to an irreversibly damaged species, or first degrading to the self-healing species and then being further degraded to an irreversibly damaged state.

\begin{figure}
\centering
\includegraphics{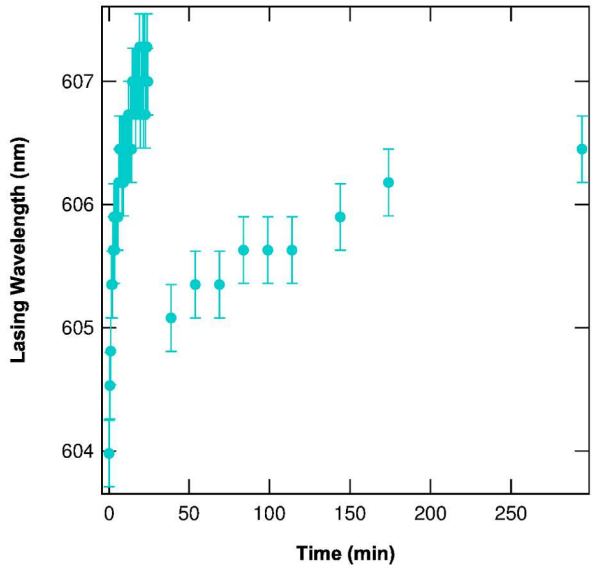}
\caption{Lasing wavelength as a function of time during decay and recovery. }
\label{fig:center}
\end{figure}

\begin{figure}
\centering
\includegraphics{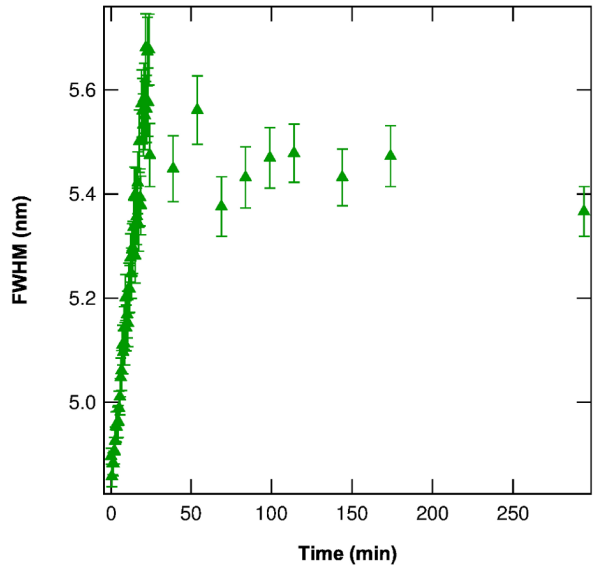}
\caption{Linewidth as a function of time during decay and recovery.}
\label{fig:linewidth}
\end{figure}

While the peak intensity fully recovers, it is inappropriate to call this process truly ``reversible'' as we find that the recovered peak is redshifted and has a broader linewidth.  Figure \ref{fig:center} shows the emission wavelength as a function of time during decay and recovery. During degradation the emission quickly redshifts, followed by a sharp blueshift once the pump beam is blocked, after which the emission slowly redshifts during recovery.  The rapid drop when the pump beam is blocked suggests that part of the redshift during decay is due to photothermal heating as removing the pump beam (and therefore photothermal heating) leads to the redshift decreasing.  This observation is consistent with previous measurements of temperature dependent RL, where increased temperature results in the RL emission redshifting \cite{Vutha06.01}. The remaining redshift (from the pristine sample) after the sample's temperature returns to the ambient temperature is most likely due to irreversible damage to the 
polymer matrix which changes the local environment of the dye and therefore the emission wavelength \cite{Cao03.01,Yi12.01,Shuzhen09.01,Anderson14.04}. Irreversible degradation of the polymer, even when the dye molecules fully recover, has been previously observed in DO11/PMMA \cite{Anderson14.01,Anderson14.03} and is most likely occurring here. Additionally, irreversible damage to the polymer can explain the observation of changes in the emission linewidth, shown in Figure \ref{fig:linewidth}, as the linewidth depends on the dye's local environment through inhomogeneous broadening \cite{Cao03.01,Yi12.01,Shuzhen09.01,Anderson14.04}.  


While irreversible polymer photodamage explains the remaining redshift after the pump is removed, as well as the change in linewidth, it does not explain the slow redshift and changing linewidth during recovery.  This observation is unexpected, but a possible explanation is that the polymer undergoes photo-oxidation and chain scission during photodegradation, with both processes are known to occur in undoped aliphatic polyurethanes \cite{Rabek95.01}, leaving the polymer in a non-equilibrium state which slowly relaxes after the pump has been removed.  

\begin{figure}
\centering
\includegraphics{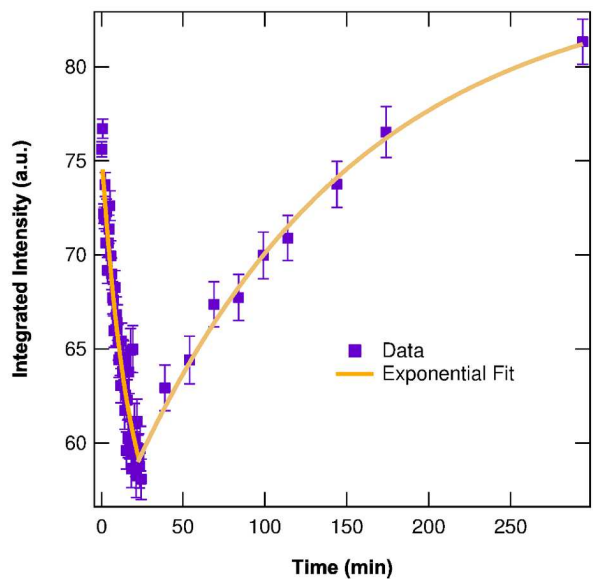}
\caption{Integrated intensity as a function of time during decay and recovery.}
\label{fig:totalint}
\end{figure}

Finally, we find a rather astonishing result when considering the integrated emission intensity as a function of time. Given the increased linewidth due to photodegradation and the observation of 100\% recovery of the peak intensity, we find that after recovery the integrated intensity is actually greater than for the pristine sample as shown in Figure \ref{fig:totalint}.  This implies that cycling the system through photodegradation and recovery actually improves the broad spectral performance of the random laser, which is important in low-coherence high-intensity imaging applications \cite{Redding11.01,Redding12.01}.   To test how much improvement is possible we perform preliminary measurements using two decay and recovery cycles and find that the first cycle leads to an increase in the integrated intensity, while subsequent cycling does not provide further enhancement.  

The observation of ``reversible'' photodegradation in R6G+ZrO$_2$/PU random lasers is a promising result for the development of robust self-healing organic-dye-based random lasers.  To better determine the underlying mechanisms of this phenomena measurements are planned using a wide range of techniques including FTIR, gas chromatography with mass spectroscopy, gel permeation chromatography, thermogravimetric analysis, Raman spectroscopy, absorbance spectroscopy, and imaging microscopy. These measurements will allow for precisely determining the relevant population dynamics, which are difficult to determine from a nonlinear process like RL.

 While the precise mechanisms are yet to be determined, they are most likely similar to the mechanisms behind DO11/PMMA's reversible photodegradation, which have been studied in detail \cite{howel02.01,howel04.01,embaye08.01,Ramini12.01,Westfall12.01,Ramini13.01,Anderson13.01,Anderson14.01,Anderson14.03}. In DO11/PMMA self-healing is proposed to be due to domains of dye molecules which are correlated via a polymer chain \cite{Ramini12.01,Ramini13.01, Anderson14.01}.  These correlated molecules interact with each other -- through the polymer chain -- such that self-healing occurs.  The exact microscopic interaction leading to self-healing is still under investigation, but possible mechanisms include domain influenced tautomerization \cite{embaye08.01,Ramini12.01,Ramini13.01}, the formation of twisted internal charge transfer (TICT) states \cite{Westfall12.01}, and photocharge ejection/recombination \cite{desau09.01,Anderson14.01}.  

One of the main predictions of the domain model developed for DO11/PMMA is that the photostability and recovery characteristics can be tuned by controlling the average domain size (which depends on the polymer host, dye concentration, temperature, and applied electric field) \cite{Ramini12.01,Ramini13.01,Anderson14.01}. If domains of correlated dye molecules are also responsible for self-healing in R6G dye-doped polymer, then by tuning the polymer properties -- as well as the nanoparticles -- it should be possible to control the photostability and recovery characteristics of a R6G+NP doped polymer random laser. This ability would allow for the development of robust photodamage resistant organic-dye-based random lasers.

This work was supported by the Defense Threat Reduction Agency, Award \# HDTRA1-13-1-0050 to Washington State University.

\pagebreak


\end{document}